\documentclass[
    ,final            
  ]
  {aipproc}

\usepackage{amsfonts}

\layoutstyle{6x9}

\begin{document}

\title{Observables and Entanglement\\ in the Two-Body System}


\author{N.L.~Harshman}{
 address={Department of Physics, American University\\ 4400 Massachusetts Ave.\ NW, Washington, DC 20016-8058}
}

\begin{abstract}
 Using the quantum two-body system as a familiar model, this talk will describe how entanglement can be used to select preferred observables for interrogating a physical system.   The symmetries and dynamics of the quantum two-body system provide a backdrop for testing the relativity of entanglement with respect to observable-induced tensor product structures.  We believe this exploration leads us to a general statement:  the physically-meaningful observable subalgebras are the ones that minimize entanglement in typical states.
\end{abstract}

\maketitle


Advances in quantum information science have offered a new set of questions with which to interrogate interacting particle systems: entanglement spectroscopy.  Physical phenomena at every scale from sub-atomic to solid state have been explored by considering the entanglement properties and dynamics of particles or quasiparticles.  In particular, the exquisite control allowed in trapped, ultracold neutral atoms and ions provides the possibility for manipulating and detecting such entanglement.  Investigating entanglement in such systems can address theoretical questions, like what is the description of a composite system that is most useful for calculation and simulation, and experimental questions, like which forms of entanglement can be controlled and exploited as a resource for quantum information processing.

This brief conference proceeding contribution summarizes some entanglement spectroscopy results that are useful for understanding a paradigmatic physical model: the quantum two-body problem.  These results showcase the connection between the notion of classical integrability and the property of quantum separability.  The two quantum bodies considered here are particles, but these results have consequences for more general bipartite divisions, including the partition between system and environment.

There are many different types of entanglement to describe and quantify in interacting particle systems.  The degrees of freedom can be continuous, discrete or mixed and are connected to the representation of space-time symmetries.  The technical challenge of how to unambiguously quantify entanglement (or more broadly, quantum correlations) in some contexts is not a solved problem, particularly for mixed states.  Further, the observables used to induce a Hilbert space representation of a state may depend on reference frame or some other perspective.  As a consequence, certain kinds of entanglement are invariant under Galilean or Poincar\'{e} transformations, others are not.  Finally, identical particles impose superselection rules on the Hilbert space that may frustrate na\"{i}ve definitions of entanglement or quantum correlation.  

Observable-induced tensor product structures is the mathematical framework used here to classify the types of entanglement, to quantify quantum correlations, and to address symmetry covariance. We review this tool in the next section, but the main idea is that one can shift the discussion from entanglement as a property of a state to entanglement as a property of a state \emph{relative} to a partition of the algebras of observables.  Sensible partitions of the observables  can be identified based on the kinematic and dynamic symmetries of an interacting system.  Applying this to the two-body problem provides support for the final argument of this article: the partition of observables that the minimizes the induced entanglement is well-suited for analyzing practical and theoretical questions within quantum few-body systems.

\section{Tensor Product Structures}

Operationally, entanglement is detected as the coherence between non-local observables, but how do we define local?  The classic scenario presents Alice and Bob, probing spin states in locations with space-like separation.  By comparing correlated expectations values for two pairs of non-orthogonal measurements, phase correlation among the expectation values, i.e.\ coherence, is revealed.  In this case, the notion of local subsystems coincides with the standard geometric notion of local.  More generally, one can define local by the subsystems themselves.  For example, though two particles in a trap are strongly interacting,the Hilbert space of the system is constructed as the tensor product of each particle's Hilbert space.  The particles are certainly not separated in space-time, but they may or may not have entangled wave functions by this `natural' notion of subsystem locality and separability.  How far can we take the subsystem definition of locality?

Zanardi's Theorem~\cite{zanardi}, gives the mathematical and physical requirements for a partition of the algebra of observables $\mathcal{A}$ into subalgebras $\{\mathcal{A}_i\}$ to induce a tensor product structure (TPS) of virtual subsystems\footnote{Defining entanglement based on observable-induced virtual subsystems is not the only extension of separability and  entanglement possible.  The term `generalized entanglement' refers to an observable-induced definition of entanglement that does not imply a partition in to subsystems or a tensor product structure~\cite{viola}.}.  Given a state space $\Phi\subset\mathcal{H}$, then the Theorem says a collection of subalgebras $\{\mathcal{A}_1,\mathcal{A}_2,\ldots \}$ of the total operator algebra $\mathcal{A}$ acting on $\mathcal{H}$ will induce a tensor product structure $\mathcal{H}=\bigotimes_i \mathcal{H}_i$ if they satisfy the following criteria: subsystem independence, completeness, and local accessibility.  The first two conditions are mathematical criteria.  Subsystem independence means that all the subalgebras commute and completeness means that the subalgebras generate the total algebra.  However, local accessibility is a physical criteria: each subalgebra must correspond to a set of controllable observables.

Zanardi's Theorem is currently proven only for Hilbert spaces realized by $\mathbb{C}^d$ with $d$ finite.  The total algebra of observables is realized as a subset of $\mathcal{A}\cong\mathbb{M}^d$, the complex $d\times d$ matrices.  Assume $d$ can be factored into positive integers as $d=\prod_i k_i$, then the subalgebras $\mathcal{A}_1 \cong \mathbb{M}^{k_1}\otimes\mathbb{I}_{k_2}\otimes\mathbb{I}_{k_3}\cdots$, $\mathcal{A}_2 \cong \mathbb{I}_{k_1}\otimes\mathbb{M}^{k_2}\otimes\mathbb{I}_{k_3}\cdots$, etc.\ induce the TPS $\mathcal{H} \cong \mathcal{H}_1 \otimes \mathcal{H}_2 \otimes \mathcal{H}_3 \cdots$.

This construction is not unique, so for every factorization of $d$, there is an equivalence class of TPSs.  Even in the simplest non-trivial case where $d=4$, alternate partitions of the observable algebra can lead to alternate evaluations of entanglement for a given state.  The Bell states $|\Phi^\pm\rangle$ and $|\Psi^\pm\rangle$ are maximally entangled with respect to commuting observable subalgebras of Alice $\mathcal{A}\cong\mathrm{span}\{{\sigma}^A_i\otimes\mathbb{I}^B\}$ and Bob $\mathcal{B}\cong\mathrm{span}\{\mathbb{I}^A\otimes\sigma_i^B\}$.  These induce a TPS $\mathcal{H}=\mathcal{H}_A\otimes\mathcal{H}_B$.  However, any unitary operator $U$ that cannot be factored into local unitaries can be used to make new subalgebras $\mathcal{P}=U\mathcal{A}U^\dag$ and $\mathcal{Q}=U\mathcal{B}U^\dag$ that induce a new TPS $\mathcal{H}=\mathcal{H}_P\otimes\mathcal{H}_Q$.  For a suitable choice of $U$, the $PQ$-type entanglement for a pure state using any measure can be anything from maximal to none~\cite{maxnone}.

For pure states, perhaps the furthest expression of this TPS-relativity is the Tailored Observables Theorem~\cite{harshman_observables_2011}: for any Hilbert space $\mathbb{C}^d$ one can construct tailored subalgebras of observables that induce a TPS from a finite basis of operators such that any known pure state can have any entanglement that is possible for any prime factorization of $d$.  The proof in \cite{harshman_observables_2011} is by construction: for a known $|\psi\rangle\in\mathbb{C}^d$ and chosen factorization of $d$, generators for the inducing subalgebras can be constructed in finite steps. 

Can these results for entanglement relativity be extended beyond pure states?  The totally mixed state is invariant under this kind of relativity of observables because $U$ commutes with the mixed state $\rho=\mathbb{I}^d$.  Mixed states are less coherent than pure states, as quantified  by $\mathrm{Tr}(\rho^2)$, and therefore have less flexibility to capture entanglement by adjusting the TPS.  For finite-dimensional Hilbert space realizations, the procedure and scope of entanglement relativity has only been explored in detail for $d=4$~\cite{thirring}.  

There are a few results for entanglement relativity in Gaussian states that can be inferred from the literature.  For example, there is always some symplectic transformation $S$ that diagonalizes the covariance matrix.  This phase space transformation can be represented as a unitary operator $U_S$ in the Hilbert space of the system. In the case of a two-mode Gaussian state, for example, the operator $U_S$ can be used to transform the canonical Heisenberg algebras $\mathcal{A}_i$ generated by $\{X_i,P_i\}$ for each mode into commuting subalgebras $U_S\mathcal{A}_i U_S^\dag$.  These new subalgebras induce a TPS with respect to which the known Gaussian state is separable.  This argument holds true for pure and mixed Gaussian states, and provides additional confirmation that Gaussian states are ``the most classical'' of quantum states.

Beyond Gaussian states, the limits of observable-induced entanglement relativity in infinite-dimensional Hilbert space realizations is an open mathematical question. How do you classify the entanglement possibilities and what observables induce them?  Given a pure or mixed state, how can one mathematically construct TPSs that minimize and maximize entanglement?  Can observables be constructed that would extract some kind of entanglement from simple, physically-accessible systems?  

\section{Symmetries of the Two Body System}

Observable subalgebras constructed to tailor the entanglement of a quantum state can be shown to exist using Zanardi's notion of induced TPSs, but these observables may not satisfy Zanardi's third condition of local accessibility.  For systems like linear quantum optics, correlated detector measurements may allow the experimentalist to reconstruct expectation values for non-local observables and realize tailored observables in the laboratory, but for interacting particle systems we cannot access the ``experimental knobs'' that would allow us to tune our observable subalgebras.  The symmetries of space and time, and the properties of the interaction and environment, determine which observables are most accessible and useful.

Consider a typical quantum two-body system, such as a proton and electron in free space, interacting via the Coulomb potential. All electron observables $\mathcal{A}_{\rm e}$ are generated by a set of free particle operators, e.g.\ $\{{\bf X}^{\rm e},{\bf P}^{\rm e},{\bf S}^{\rm e}\}$. A similar set generates $\mathcal{A}_{\rm p}$ for the proton.  The vector operators for position ${\bf X}$, momentum ${\bf P}$ and intrinsic spin ${\bf S}$ are themselves constructed from the Lie algebra of Galilean spacetime transformations.  These are natural observables when the electron and proton are far apart and weakly-interacting, but for bound states these observables are less accessible.  

The total algebra of observables $\mathcal{A}$ is the direct sum $\mathcal{A}_{\rm e}\oplus\mathcal{A}_{\rm p}$ and the tensor product structure induced is $\mathcal{H}=\mathcal{H}_{\rm e}\otimes\mathcal{H}_{\rm p}$.  Because of this construction, Galilean transformations are represented as local unitary operators with respect to this TPS and therefore interparticle entanglement is a Galilean invariant for pure states and mixed states.  In elastic scattering, one generally assumes the asymptotic in-state has no interparticle entanglement, and therefore scattering can only increase it.  One can show in a variety of model systems that almost every scattering interaction leads to an increase in entanglement for almost every unentangled in-state~\cite{scattering}.  For bounds states, interparticle entanglement is also the norm, even for the simple system of coupled harmonic oscillators~\cite{osc}.

Each particle Hilbert space is realized as square-integrable/square-summable functions on the spectra of a complete set of observables, for example the choice $\{{\bf X}^{\rm e},{S}_z^{\rm e}\}$ gives us the realization $\mathcal{H}_{\rm e}=\mathcal{H}^{\bf x}_{\rm e}\otimes\mathcal{H}^s_{\rm e} = \mathrm{L}^2(\mathbb{R}^3)\otimes\mathbb{C}^2$.  More generally, the subalgebras $\mathcal{A}^{\bf x}_{\rm e}$ and $\mathcal{A}^{\rm s}_{\rm e}$ generated by $\{{\bf X}^{\rm e},{\bf P}^{\rm e}\}$ and $\{{\bf S}^{\rm e}\}$ are complete and commuting, and they can be used to defined the tensor product structure for intraparticle entanglement between motional and spin degrees of freedom for a single particle.  For the electron, the maximum entanglement with respect to this TPS is the same as for a TPS of two qubits~\cite{continuous}, and it is also a Galilean invariant~\cite{osid}.

The Hamiltonian $H$ is an operator in $\mathcal{A}$ that cannot be expressed as the sum of an electron operator in $\mathcal{A}_{\rm e}$ and and a proton operator in $\mathcal{A}_{\rm p}$.  However, we know that there is a special symplectic transformation $S$ compatible with the overall Galilean symmetry that separates a Hamiltonian with a central interaction.  In phase space, the transformation $S$ is a linear map that takes the canonical coordinates $({\bf x}^{\rm e},{\bf x}^{\rm p},{\bf p}^{\rm e},{\bf p}^{\rm p})$ into the center-of-mass and relative coordinates  $({\bf x}^{\rm c},{\bf x}^{\rm r},{\bf p}^{\rm c},{\bf p}^{\rm r})$.

The unitary representation of this transformation $U_S$ leaves invariant the total algebra $\mathcal{A}=U_S \mathcal{A} U_S^\dag$ and each particle's spin subalgebra $\mathcal{A}^{\rm s}_{\rm e}$ and $\mathcal{A}^{\rm s}_{\rm p}$.  The representation $U_S$ linearly transforms the generators of the motional subalgebras $\mathcal{A}^{\bf x}_{\rm e}$ and $\mathcal{A}^{\bf x}_{\rm p}$ into the generators of the center-of mass $\mathcal{A}^{\bf x}_{\rm c}$ and relative $\mathcal{A}^{\bf x}_{\rm r}$ motional observables.  The decomposition of $\mathcal{A}$ into $\mathcal{A}^{\bf x}_{\rm c}\oplus\mathcal{A}^{\bf x}_{\rm r}\oplus \mathcal{A}^{\rm s}_{\rm e}\oplus\mathcal{A}^{\rm s}_{\rm p}$ provides another Galilean invariant TPS.

The transformed central-force two-body Hamiltonian $U_S {H} U_S^\dag$ decomposes into center-of-mass Hamiltonian ${H}^{\rm c}\in\mathcal{A}^{\bf x}_{\rm c}$, plus the relative Hamiltonian ${H}^{\rm r}\in\mathcal{A}^{\bf x}_{\rm r}$. This separability is a quantum manifestation of classical two-body solvability; Galilean symmetry applied to phase space provides enough independent Casimir invariants to integrate the motion.
For the center-of-mass (or external) Hilbert space $\mathcal{H}_{\rm ext}=\mathcal{H}^{\bf x}_{\rm c}$, a complete set of commuting observables is $\{{\bf X}^{\rm c}\}$ and wave functions are square-integrable functions on their spectrum $\mathbb{R}^3$.  

The relative Hamiltonian $\hat{H}^{\rm r}$ for a central, spin-independent interaction acts only on $\mathcal{H}_{\rm rel}=\mathcal{H}^{\bf x}_{\rm r}$.  One could realize this Hilbert space as functions on the spectrum of the complete commuting set $\{{\bf X}_{\rm r}\}$, but these operators do not commute with the relative Hamiltonian.  A typical set would be $\{{H}_{\rm r},\hat{L}^2,\hat{L}_z\}$, and the Hilbert space spectrum of ${H}^{\rm r}$ decomposes into a discrete, accumulating set of bound states and a continuum of scattering states.  More generally, a spin-dependent non-central Hamiltonian would be an operator in the internal subalgebra $\mathcal{A}_{\rm int}=\mathcal{A}^{\bf x}_{\rm r}\oplus\mathcal{A}^{\rm s}_{\rm e}\oplus\mathcal{A}^{\rm s}_{\rm p}$.  

The induced TPS $\mathcal{H}=\mathcal{H}_{\rm ext}\otimes\mathcal{H}_{\rm int}$ is not just Galilean invariant, but it is dynamically invariant because the Hamiltonian exponentiates to a time evolution operator that is local with respect to the internal-external subalgebras~\cite{prl}.  For a hydrogen atom, the center-of-mass motion will remain unentangled with the internal atomic states as long as there are no external fields that couple them.  For example, a non-uniform external (classical) electrostatic or magnetostatic field will cause a dipole force that depends on the relative state, possibly leading leading to changes in the internal-external entanglement.  For some external fields and some states, e.g.\ harmonically-trapped, equal mass particles, the internal-external entanglement will also be time-invariant~\cite{osc}.

This example shows that the entanglement with respect to TPSs induced by certain observable subalgebras can be more useful for describing the quantum two-body system than others.  Depending on the space-time symmetries and the particular system dynamics, different types of entanglement are invariant. In isolated hydrogen atoms, we expect lower internal-external entanglement than interparticle entanglement for typical states.  This can be extended to harmonic oscillators, whose Gaussian ground states are unentangled in the mode coordinates, but highly entangled in the particle coordinates.  In other words, using the `right' observables minimizes the entanglement.  They shift the coherence and quantum correlations to `local' observables.  Whether this notion can be used to find efficient schemes for approximating non-integrable systems, for example the three-body system, seems like an interesting topic for future research.





\bibliographystyle{aipproc}   



\end{document}